\documentclass[journal=jacsat,manuscript=article]{achemso}

\usepackage{chemformula} 
\usepackage[T1]{fontenc} 

\newcommand{\sio}{SiO$_2$ }
\usepackage{color,soul, cancel}
\usepackage{ulem}

\usepackage{textcomp}
\setkeys{acs}{maxauthors=10, etalmode=truncate}

\author{Viktoriia Rutckaia}
\email{viktoriia.rutckaia@physik.uni-halle.de}
\affiliation[Martin-Luther-University]
{Centre for Innovation Competence SiLi-nano\textsuperscript{\textregistered}, Martin-Luther-University of Halle-Wittenberg, Karl-Freiherr-von-Fritsch-Str. 3, 06120 Halle (Saale), Germany}

\author{Frank Heyroth}
\affiliation[Martin-Luther-University]
{Interdisciplinary center of material science, Martin-Luther-University of Halle-Wittenberg, Heinrich-Damerow-Str. 4, 06120 Halle (Saale), Germany}

\author{Georg Schmidt}
\affiliation[Martin-Luther-University]
{Institute of Physics, Martin-Luther-University of Halle-Wittenberg, von-Danckelmann-Platz 3, 06120 Halle (Saale), Germany}

\author{Alexey Novikov}
\affiliation[IPM  RAS]
{Institute  for  Physics  of Microstructures  of  the Russian  Academy of  Sciences, Academicheskaya Str. 7, Nizhny Novgorod 603950, Russia}
\alsoaffiliation[]{Lobachevsky University, Gagarin av. 23, Nizhny Novgorod
603950, Russia}

\author{Mikhail Shaleev}
\affiliation[IPM  RAS]
{Institute  for  Physics  of Microstructures  of  the Russian  Academy of  Sciences, Academicheskaya Str. 7, Nizhny Novgorod 603950, Russia}

\author{Roman Savelev}
\affiliation[ITMO University]
{Department of Nanophotonics and Metamaterials, ITMO University, St. Petersburg 197101, Russia}

\author{Joerg Schilling}
\affiliation[ZIK]
{Centre for Innovation Competence SiLi-nano\textsuperscript{\textregistered}, Martin-Luther-University of Halle-Wittenberg, Karl-Freiherr-von-Fritsch-Str. 3, 06120 Halle (Saale), Germany}

\author{Mihail Petrov}
\affiliation[ITMO University]
{Department of Nanophotonics and Metamaterials, ITMO University, St. Petersburg 197101, Russia}

\title[Luminescence Enhancement in One-Dimensional Mie-Resonant Arrays]
  {Luminescence Enhancement in One-Dimensional Mie-Resonant Arrays}


\begin{document}



\begin{abstract}
  In this paper, we demonstrate the  infrared photoluminescence emission from Ge(Si) quantum dots enhanced with collective Mie modes of silicon nanopillars. We show that the excitation of band edge dipolar modes of a linear  nanopillar array results in strong reshaping of the photoluminescence spectra. Among other collective modes, the magnetic dipolar mode with the polarization along the array axis contributes the most to the emission spectrum exhibiting an experimentally measured Q-factor of around 500 for an array of 11 pillars. The  results belong to the first experimental evidence of light emission enhancement of quantum emitters applying collective Mie resonances and therefore represent an important contribution to the new field of active all-dielectric meta-optics. 
  
\end{abstract}

\section{Introduction}

All-dielectric resonant nanophotonic structures have emerged recently as a powerful platform for light manipulation at the sub-wavelength scale. The tunable electric and magnetic  Mie resonances of high refractive index nanostructures represent the basis of a variety of optical effects such as manipulation of the scattering patterns with dielectric nanoantennas~\cite{Fu2013, Staude2013, Sinev2017a,Shibanuma2017a}, tailoring of reflection and transmission~\cite{Moitra2015, Babicheva2017} as well as control of the wavefront~\cite{Kamali2018} with dielectric metasurfaces. Compared to plasmonic materials, dielectrics and semiconductors possess low and often negligible Ohmic losses which leads to higher efficiencies of nanophotonic devices~\cite{Kruk2017}. For instance, a significant progress in the field of nonlinear light emission from single nanoparticles~\cite{Carletti2015, Xu2020, Smirnova2016}, nanoantennas~\cite{Capretti2017a, Cambiasso2017a} and metasurfaces~\cite{Wang2018, Sain2019} has been demonstrated using dielectric materials.

Although all-dielectric Mie-resonant systems have been initially explored only in the passive domain, recently their use was extended to the field of active nanophotonics~\cite{Bidault2019,Staude2019,Liu2018, Pero2020}, where they are applied in the search for compact light-emitting devices. Nevertheless, the active systems with integrated quantum emitters are still in the very early stage of development. Similar as in active plasmonics~\cite{Jiang2018}, the coupling of quantum emitters such as colloidal quantum dots and dye molecules with compact all-dielectric nanostructures has also been studied~\cite{Regmi2016, Sanz-Paz2018, Cambiasso2017a, Vaskin2019}. However, for emitters placed on the surface of the nanostructures the coupling with Mie modes, which are mainly localized inside the resonator, is not very efficient. Therefore an alternative approach based on the integration of active media such as quantum dots inside the Mie resonant structure fabricated e.g. via  epitaxial deposition processes has also been suggested recently~\cite{Rutckaia2017,Capretti2017a}. 

The coupling strength between quantum emitters and resonant optical modes, defined by the ratio $Q/V$ of quality factor $Q$ and mode volume $V$, is limited in the case of sub-wavelength Mie nanostructures because of their strong radiative losses. In order to overcome this fundamental limitation, one needs to suppress the far-field radiation through a proper engineering of the radiative losses of the resonant modes. In single nanoresonators, it was recently suggested to exploit the hybridization of two modes~\cite{Rybin2017, Koshelev2019} with the aim to cancel a large part of the far-field radiation produced by certain multipole moments. That has been implemented and record high efficiency of second-harmonic radiation from single dielectric nanoparticles~\cite{Koshelev2020} and even lasing regimes of light emission~\cite{Ha2018, Mylnikov} was observed.

Another approach adopted in this work is based on the interaction of dipole modes in nanoresonator oligomer ensembles~\cite{Bulgakov2019-2, Bulgakov2018} or one-dimensional array systems~\cite{Asenjo-Garcia2017,Kornovan2016} with the  formation of collective discrete-dipole states exhibiting low radiative losses. Such systems have been extensively studied (both, theoretically~\cite{Savelev2014} and experimentally~\cite{Savelev2015,Bakker2017a}) in the context of light-guiding. In addition a strong  Purcell enhancement up to a factor of 70 in  finite size structures consisting of 10 particles was experimentally shown in the microwave region~\cite{Krasnok2016}.
  
In this paper, we consider an active photonic cavity based on a chain of coupled nanopillar resonators with embedded Ge(Si) quantum dots (QD). We demonstrate for the first time the coupling of the QDs to the collective Mie-modes of the cavity, which exhibit increasing Q-factors for a rising number of elements in the chain. We propose a bottom-up approach to design Mie-mode-based one-dimensional (1D) photonic crystal chains for manipulating the photoluminescence emission from the integrated quantum sources. 
The fabrication of the structures was done  using Si based standard semiconductor processes and the measured luminescence spectrum of the Ge(Si) quantum dots covered several telecom bands from 1200 to 1700 nm, which makes the proposed structures compatible with modern CMOS technology and opens a perspective for practical applications.

\section{Results and discussion}

\subsection{Isolated resonator}

\begin{figure}
  \includegraphics{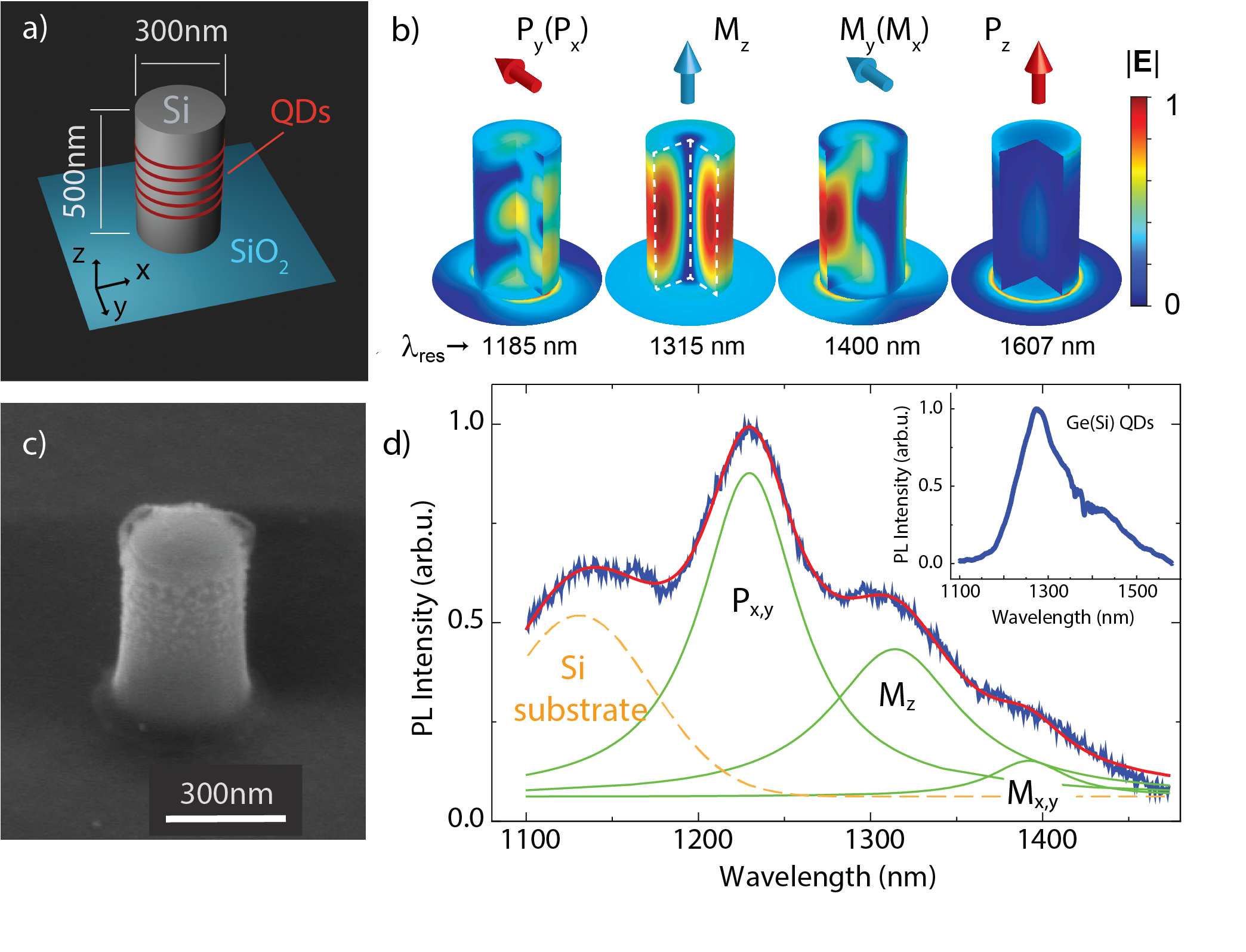}
\caption{a) Schematic of the pillar with embedded layers of QDs, the resonator stands on the layer of \sio; b) Normalized electric field distribution of electric and magnetic dipole modes (color map). Directions of the electric and magnetic dipoles are shown with red and blue arrows, respectively. c) SEM image of an isolated Si nanopillar with Ge(Si) QDs.  d) $\mu$PL spectrum of the pillar (blue line) and its fitting (red line) with three Mie-resonances (green lines) and the signal from the underlying substrate (yellow dashed line); inset shows $\mu$PL signal of Ge(Si) QDs embedded in an unstructured silicon layer.}\label{Fig1}
\end{figure}

We start with identifying the modal content of a single silicon pillar that serves as a building block of the 1D resonant array. In Fig.~\ref{Fig1}a), the schematic of an isolated circular Si pillar containing several embedded layers of Ge(Si) QDs is shown. It stands on the 3 $\mu$m \sio layer backed by the  Si handling substrate. For the chosen height $h=500$ nm and diameter $d=300$ nm of the pillar the electric and magnetic dipole modes of lowest order occur in the 1200-1400 nm spectral range~\cite{Rutckaia2017} overlapping with the luminescence range of the Ge(Si) quantum dots. The higher order (quadrupole) modes of the pillars lie in the shorter wavelength region outside the QD luminescence spectral range, therefore, from what follows we focus only on the dipole modes. The electric field distributions for the dipole modes calculated with the commercially available finite element modelling package COMSOL is shown in Fig.~\ref{Fig1}b) in the order of increasing resonant wavelength. The direction and type of the corresponding dipole moments are shown with arrows. Due to the rotational symmetry, all horizontally oriented dipoles are equivalent causing the degeneracy of $x$- and $y$-oriented dipoles (in Fig.~\ref{Fig1}b) only the $y$-dipole field distribution is shown). The high aspect ratio of the pillar causes the spectral separation of in-plane ($x$- and $y$-) and out-of-plane ($z$-) oriented modes.  One should notice that the \sio layer under the pillar does not destroy the resonances~\cite{Groep2013} due to its relatively low refractive index.

If a quantum emitter such as a QD is placed inside the pillar, its spontaneous emission  can be suppressed or enhanced due to the Purcell effect~\cite{Purcell1946}. A detailed investigation of the QD coupling with the Mie modes in a cylindrical pillar structure was reported in Refs.~\citenum{Rutckaia2017,Rocco2017}. In order to test the coupling properties experimentally,  single nanopillars comprising 12 layers of Ge(Si) QDs were fabricated; Fig.~\ref{Fig1}c) shows a scanning electron microscope (SEM) image of such a nanopillar. In the experiment, the pillars were excited by a focused 532 nm continuous-wave laser and the micro-photoluminescence ($\mu$PL) signal from the structure was collected by an objective lens placed above the sample. The Purcell effect along with the specific radiation patterns of the resonance modes result in the reshaping of the measured luminescence spectrum leading to increase of the detected emission at the resonance wavelengths.  The measured $\mu$PL spectrum of the nanopillar, shown in Fig.~\ref{Fig1}d) with the blue curve, was normalized to the luminescence spectrum of the unprocessed part of the sample with Ge(Si) QDs (shown in the inset) and fitted with four lorentzians (red curve) in order to elucidate the impact of the Mie resonances on the emission spectrum. The contribution around 1150 nm originates from the Si emission of the  substrate  beneath the \sio layer (yellow dashed). The other three peaks (green) correspond to the dipole resonances shown in Fig.~\ref{Fig1}b) confirming the coupling of Ge(Si) QDs to the Mie-modes. The shift of the experimental peak positions in comparison with the theoretically calculated positions of the resonances (electric dipole modes  $P_{x(y)}$ and $P_{z}$:  1185 nm  and 1607 nm, respectively; magnetic dipole modes  $M_{x(y)}$ and $M_{z}$: 1400 nm and 1315 nm, respectively) can be attributed to the fabrication intolerance, e.g., the slightly conical shape of the resonator, surface roughness and an effective local increase of the refractive index of the pillar due to the germanium contribution from the QDs.

\subsection{Chains of pillars}

\begin{figure}
  \includegraphics{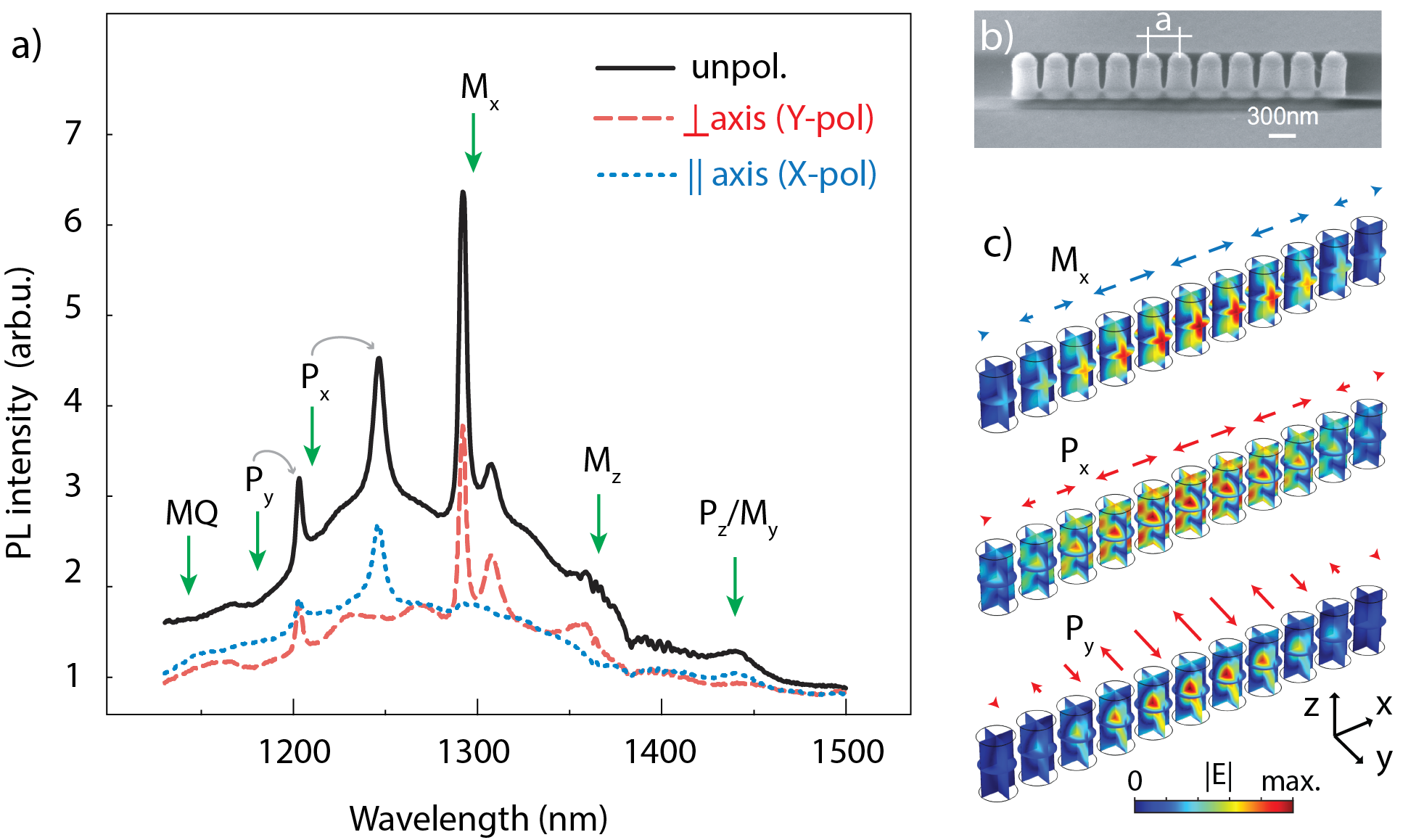}
\caption{a) PL spectra of a  Mie-resonator chain consisting of 11-pillars with embedded QDs: unpolarized (black solid line), $y$-polarized - perpendicular to the chain axis (red dashed line) and $x$-polarized - along the chain axis (blue dotted line). Green arrows mark calculated resonant wavelength of the nanopillar array; b) SEM image of the resonator chain with the height of each pillar 500 nm, diameter 300 nm, gaps between neighbouring elements 70 nm;  c) Normalized electric field distribution of coupled electric and magnetic dipole modes (color map), directions and relative magnitudes of the magnetic and electric dipole moments induced in the pillars are shown with blue and red arrows, respectively.
}\label{Fig2}
\end{figure}

In order to observe the collective Mie resonant states, we have fabricated a 1D array of the nanopillars with the same radius and height and a period $a= 370$ nm (i.e. 70 nm gap between the pillars). Fig.~\ref{Fig2}a) (black) shows the measured $\mu$PL-spectrum of the array consisting of 11 pillars, with the SEM image shown in Fig.~\ref{Fig2}b). In the spectrum, there are three sharp peaks (around 1200, 1240 and 1290 nm) that correspond to the excitation of the collective resonant states as well as several less pronounced peaks. To further characterize these states, we carried out polarization-resolved $\mu$PL measurements by placing a polarizer in the collection line 
and aligning it either along the chain axis ($x $-axis) or perpendicular to it ($y $-axis). The resulting spectra are shown with red dashed and blue dotted curves in Fig.~\ref{Fig2}a). One can see that  the  peaks at 1200 nm and 1290 nm exhibit a strongly $y$-polarised radiation pattern, while the peak at 1240 nm shows a dominant polarisation along the  $x$-direction.

We also checked the polarization-resolved $\mu$PL of the unstructured substrate to confirm that an ensemble of pristine QDs emits unpolarized light (see Supporting Information).

In order to classify the resonant peaks in the spectra, we have performed a numerical modelling of the 1D array. The modes of the finite arrays represent high-Q collective states which emerge due to the coupling between the dipole modes of individual nanopillars. By analysing the field distribution and local polarisation of the modes of the chain, they can be attributed to the particular dipolar modes of the individual nanopillars they originate from. The simulated spectral positions of the highest Q-modes are marked with green arrows in Fig.~\ref{Fig2}a) and the corresponding electric field intensity distributions for 3 particular resonances are shown in Fig.~\ref{Fig2}c). The notation of the chain modes indicates the types of resonances of the individual nanopillars from which they are constructed. In Fig.~\ref{Fig2}c), the orientations of dominating individual magnetic and electric dipole moments in each pillar in the chain are indicated by the blue and red arrows, respectively, while the arrows' length qualitatively reflects the magnitude of the individual dipole moments according to the performed calculations. From this plot one can observe that the phase of the electromagnetic field alternates by $\pi$ in the neighbouring nanopillars, which is essential in the suppression of the radiative losses.

The polarisation of the collective dipole modes provides the opportunity to identify them reliably in the experimentally observed polarisation-resolved luminescence spectra. 
The modes corresponding to excitation of $M_x$ or $P_y$ ($P_x$ or $M_y$) dipole moments contain major electric (magnetic) field components in $y$ ($x$) direction and therefore the emission is expected to be predominantly polarized along the  $y$ ($x$) axis. In addition to this, the $P_z$ and $M_z$ modes only weakly contribute to the measured $\mu$PL signal as they do not radiate in the vertical direction, and only a small part of the off-vertical radiation is collected by the microscope objective. The calculated frequencies of the collective resonances along with their polarization agree well  with the measured $\mu$PL-spectra, as seen in Fig.~\ref{Fig2}a). Like for the single pillar resonances we note that the slight discrepancies between the spectral positions of theoretically identified modes and the experimentally measured peaks are related to the fabrication imperfections and increased refractive index due to embedded Ge(Si) QDs layers. We indicate the shifted experimental peaks and the corresponding theoretically calculated resonance positions with gray arrows. Additional simulations taking  into account the non-ideal shape of the fabricated nanopillars and  influence of Ge(Si) QDs layer on the positions of the resonances are available in Supporting Information section.

\begin{figure}[t]
  \includegraphics[width = \textwidth]{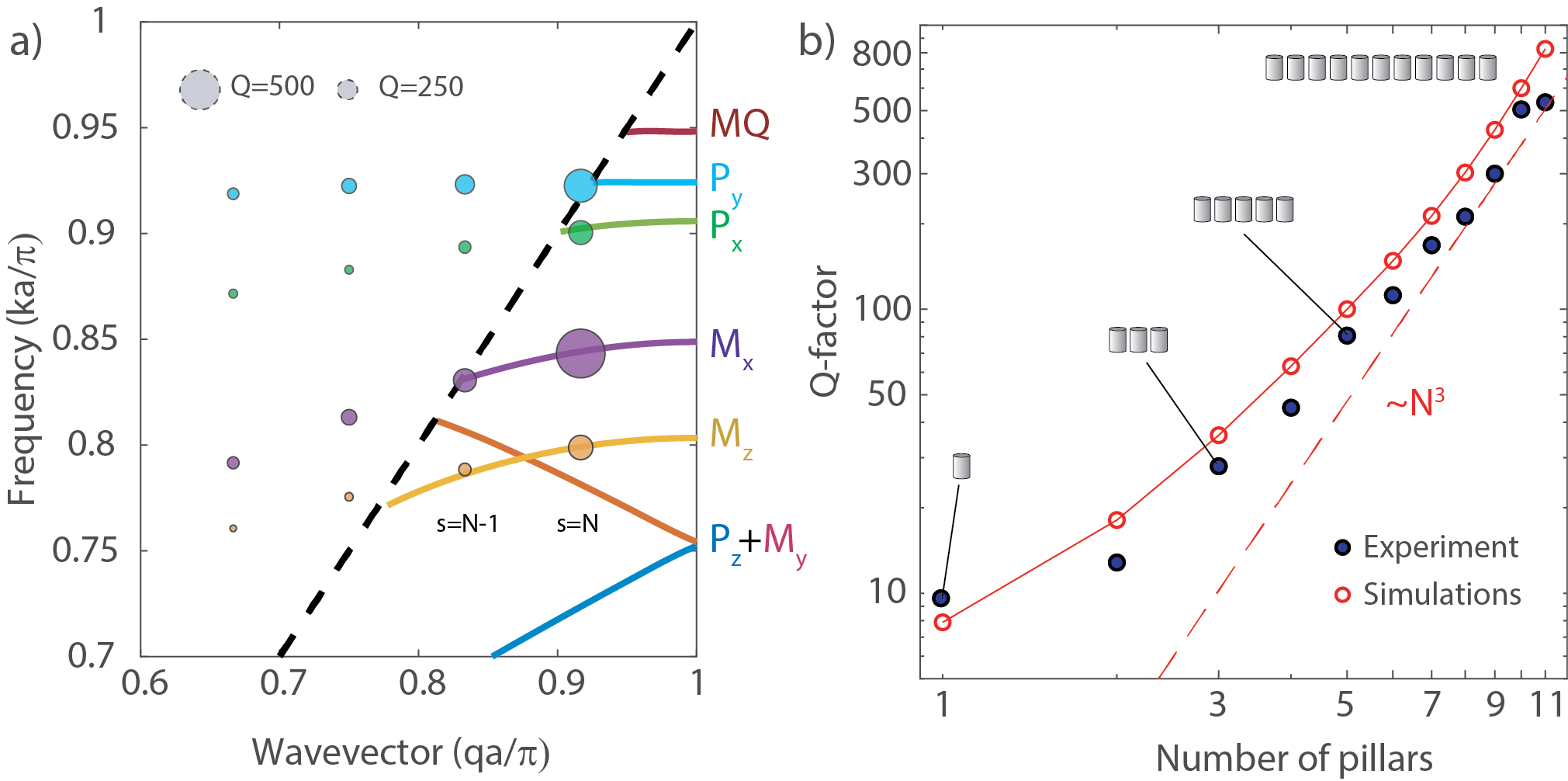}
\caption{a) Dispersion diagram (colored solid lines) of the infinite discrete chain for electric dipole ($P_x, P_y, P_z$), magnetic dipole ($M_x, M_y, M_z$) and magnetic quadrupole modes ($MQ$). The circles indicate eigenmodes of a finite chain of 11 nanopillars. The sizes of the circles reflect the Q-factors of the modes. b) Dependence of the Q-factor of the dominating collective $M_x$-resonance on the length of the chain. With the increasing number of pillars the Q-factor rises in a super-linear way. Dark blue points mark experimentally observed values. The red solid line indicates Q-factors obtained numerically. 
}\label{Fig3}
\end{figure}

The physical picture underlying the luminescence enhancement of the modes with alternating $\pi$ phase shift can be immediately understood by considering the transition from the excitation in an infinite Mie-resonant chain to collective resonances in a finite chain. For an infinite one-dimensional array of pillars, the hybridized modes form Bloch states where the electric ($E$) and magnetic ($H$) fields are described by the product of a periodic function $u(x)$ and the phase factor $\exp(iqx)$, where $q$ is the Bloch wavenumber. The periodic function $u(x)$ typically corresponds to the field pattern of one of the resonances in a single pillar, e.g., the $M_x$ Mie-resonance, while the $\exp(iqx)$ factor describes the phase of the field oscillation in each element. In the dispersion diagram $\omega(q)$, these Bloch-states appear as continuous photonic bands, where each band corresponds to a specific Mie-resonance or in some cases their combination when they are spectrally overlapped. Fig.~\ref{Fig3}a) shows the calculated dispersion curves for an infinite linear array of Si-pillars as well as the eigenmodes of the finite-length Mie-chains. When the wavevector $q$ is large the Mie-modes of the infinite array (indicated by the coloured continuous curves) lie below the light line and therefore exhibit zero radiative losses. Thus, such a resonant pillar array forms a discrete waveguiding system well-known in all-dielectric photonics~\cite{Savelev2014-2,Savelev2015,Bakker2017a} and in plasmonics~\cite{Koenderink2006,Petrov2015}. Above the light line, the Mie-modes of the infinite array efficiently radiate into the surrounding and therefore are characterized by low Q factors (not shown in the figure). Based on the analysis of the field profile and multipolar decomposition we have identified the types of Mie-bands and the associated particular dipole modes (see Fig.~\ref{Fig3}a), right side). Note that not only the same type of the dipole modes of individual nanopillars can couple to each other when combined in a chain. Dipole modes possessing the same type of symmetry with respect to the plane $y=0$ can couple to each other if they are spectrally overlapped. In the considered structure only $P_z$ and $M_y$ resonances have close enough eigenfrequencies to have substantial interaction, and therefore they form the lowest two branches.

In a finite chain consisting of $N$ pillars, for each of the Mie-band of the infinite chain there exists a finite number ($N$) of hybridized modes, that form discrete representations of these continuous bands. In Fig.~\ref{Fig3}a), we show them as circles with the diameters that reflect the corresponding Q-factors (larger circle indicates larger Q-factor). 
Their positions in the reciprocal space can be approximately determined by performing the discrete Fourier transform of the dipole moments distribution. The dipole moments in the mode with the largest Q-factor typically oscillate out-of-phase and therefore such mode has a corresponding wavenumber $q \approx \pi N/(a(N+1))$~\cite{Weber2004, Figotin2005}.
The out-of-phase oscillations of closely spaced Mie-dipoles lead to the destructive interference of their far fields. For an infinite array, the interference is perfect resulting in zero radiation losses (as one can expect since the modes near the Brillouin-zone edge fall below the light line). For a finite chain the cancellation is incomplete resulting in the non-zero coupling of the chain modes to the free space modes causing slight radiation losses. However, with an increasing number of pillars in the chain, the system approaches the infinite array and the radiative losses decrease. Theory predicts that the radiative Q-factor has a $N^3$ dependence~\cite{Blaustein2007, Polishchuk2017,Bulgakov2019,Sadrieva2019} in the limit of large $N$ showing the potential to achieve even higher Q-factors for longer chains.

We have studied the evolution of the collective Mie-resonances for different chain lengths by fabricating 10 chains (whose number of the nanopillars varied from 2 to 11) and measuring their $\mu$PL spectra. All spectra showed pronounced collective resonances which appeared considerably sharper than in the peaks of the single pillar. We have extracted the Q-factors of the resonant modes by fitting Lorentzians to the corresponding peaks in the PL signal. As an example the dependence of the Q-factor of the dominating  $M_x$ mode with increasing number of pillars is shown in Fig~\ref{Fig3}b) (other modes have similar dependence). An overall super-linear increase of the Q-factor is observed with an increasing number of pillars reaching the measured value  of $Q\approx 500$. One can note that starting from 8-9 nanopillars, the Q-factor dependence becomes cubic in full agreement with the theory.  The numerical simulation  (solid red line) agrees well with the measured results predicting however higher values of Q-factors.

\subsection{Other one-dimensional edge mode structures   }

\begin{figure}[t]
\includegraphics{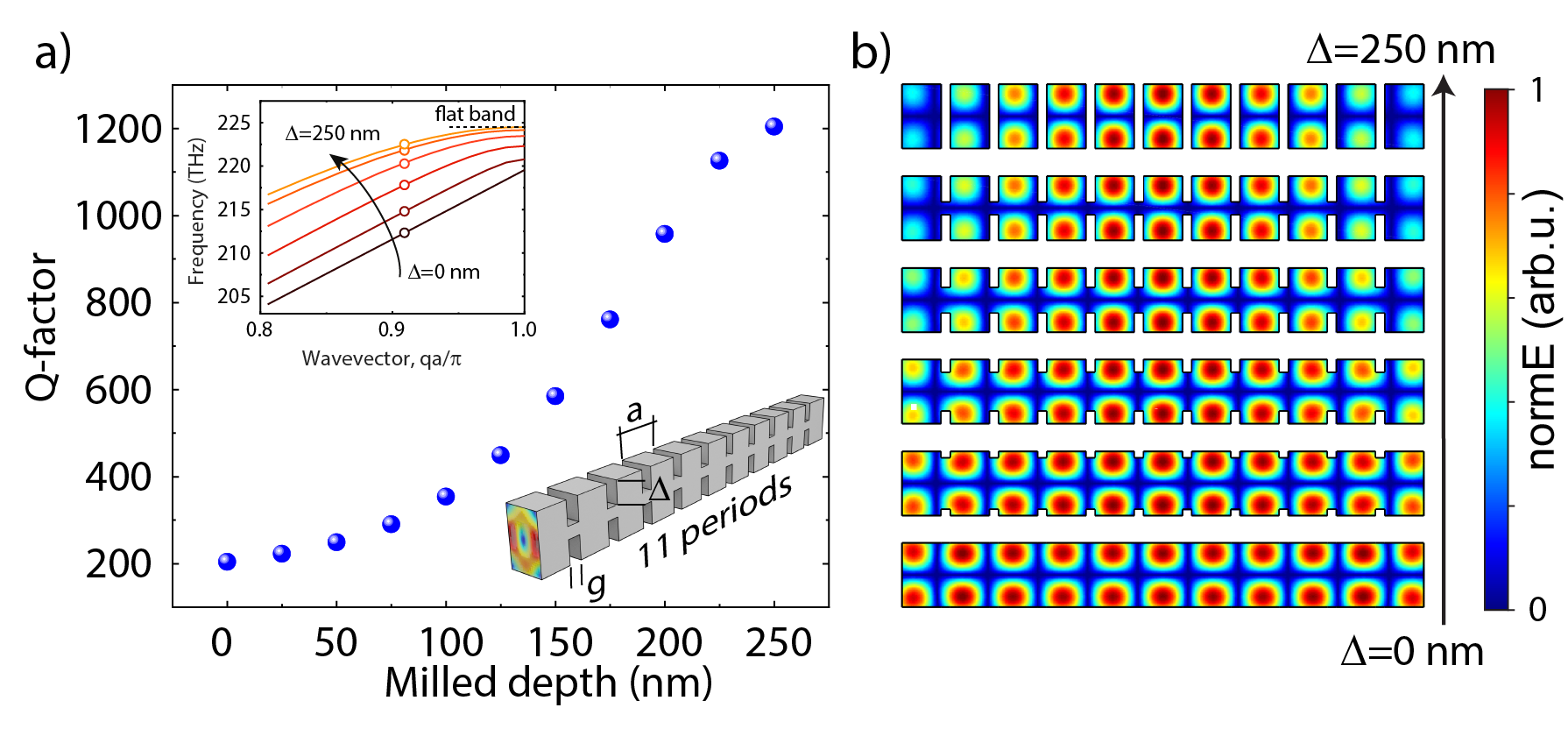}
\caption{a) Q-factor of the mode with the smallest radiation losses of the finite-size bar structure with ten grooves as a function of the depth of the grooves $\Delta$. The width of the groove is $h=70$ nm and period is $a=370$ nm, which corresponds to the parameters of nanopillar array shown in Fig.~\ref{Fig2}. The inset shows the dispersion in the corresponding infinite ridge waveguiding structure. The modes of the finite bar are marked with circles. b) Cross sections of the field intensity profiles for high-Q modes in the finite bar structure with grooves. }\label{Fig4}
\end{figure}

To  elucidate the origin of the higher Q-factors for the longer 1D Mie-resonant arrays a more detailed study of the collective mode profile was performed. Instead of adding more and more individual resonators to form a long chain, the same chain of Mie resonators can also be approached by introducing a periodic array of grooves with increasing depth into an originally unstructured bar of the same final length. In this way one can trace the evolution of the final collective Mie resonant modes from the Fabry-Perot-type (FP) modes of the originally unstructured dielectric bar (Fig.~\ref{Fig4}). The finite rectangular  dielectric bar of 300 nm width and 500 nm height, which corresponds to the geometrical parameters of cylindrical  nanopillars, supports a $E^y_{21}$ Fabry-Perot type mode. This mode has a strong $H_x$ component of the magnetic field  matching the mode profile of the $M_x$ mode of the nanopillar array. 
Adding the grooves of increasing depth, a smooth transition from the extended FP mode to a coupled discrete dipole Bloch-mode in the isolated nanopillar chain is achieved. The underlying periodicity of the FP standing wave pattern remains the same  also in the collective Mie-mode of the nanopillar chain leading there to the discussed out-of-phase oscillation in neighbouring individual pillars. However a remarkable rise of the Q-factor is observed  along with the  transformation from the unstructured bar to the chain of fully isolated resonators Fig.~\ref{Fig4}a). The discussed out-of phase oscillation can therefore not explain the high Q-factors in longer chains alone. A careful inspection of Fig.~\ref{Fig4}b) also reveals that, with increasing groove depth, an additional envelope function appears, which governs the local mode amplitude. The envelope resembles a half-cosine wave which extends over the whole chain leading to a maximum mode amplitude at the centre and minima close to the two ends of the chain. This is in stark contrast to the constant amplitude across the whole unstructured bar, where the mode appears to be "chopped off" at the ends. The extension and the smoothing of the intensity profile towards the ends of the chain are important additional factors supporting the reduction of the radiation losses from the overall chain~\cite{Polishchuk2017, Poddubny2020}. This was already observed for photonic crystal cavities~\cite{akahane_high-_2003, Vukovic2001, Englund:05}, where up- and downward directed radiation losses depend on the overlap integral of the overall mode with the plane waves travelling into the far field above and below the chain. This integral basically represents the spatial Fourier transform of the overall mode pattern. Although the position of the  major Fourier peak in k-space is determined by the underlying periodicity of the anti-nodes, the spectral spread of the Fourier peak is governed by the envelope. An extended smooth envelope in real space leads then to a highly localised Fourier transform for k-values mainly below the light line, whose corresponding waves can not travel into the far field resulting in low radiation losses and a high Q-factor. Sharply localised mode profiles on the other hand with abrupt transitions (e.g. "chopped off" modes)  exhibit Fourier transforms with much wider spreads in k-space including Fourier components with k-values above the light line, which result in larger radiation losses and lower Q-factors. 
Especially in photonic crystal nanobeam structures extensive efforts were therefore undertaken to achieve ultra high Q-cavities by shifting and adjusting pore sizes aiming for a smooth transition between the cavity mode profile and the surrounding photonic crystal mirror region~\cite{Sauvan:05, Schriever:10}. 

Indeed the photonic crystal nanobeam structures~\cite{Foresi1997, Notomi2008} are the closest to be compared to our structures as they possess similar geometry and footprint.  Photonic crystal cavities and waveguides on the other hand require a considerably larger surrounding area with a periodically structured refractive index. The nanobeam structures have already shown their potential for various applications in optics and photonics~\cite{Kim2011} as well as for sensing applications~\cite{Zhang2015,Yang2020}. Other studies show that the band edge states in the finite length nanobeam structures with  similar geometry provides  Q-factors of several hundreds~\cite{Burr2013, Kim2011} which is comparable to the result reported in this work.  

However in contrast to conventional photonic crystal systems, we propose the Bragg edge states formation through engineering of the collective Mie states, which can be considered as a bottom-up approach to resonant structure design. The Mie-cavity here acts as a building block for more complex structures, which also opens a new perspective for Q-factor optimization through higher order multipoles engineering~\cite{Kornovan2020, Liu2020}. Going beyond the dipole modes reported in this paper, the high-Q single Mie resonators operating in the quasi-BIC regime~\cite{Rybin2017,Huang2019, Bulgakov2020} and reaching the Q-factors of around hundred attracted a lot of attention recently. In the future these quasi-BIC-Mie resonators could be utilized as unit cells in novel collective Bragg resonant systems reaching  even higher Q-factors by combining the radiative loss cancellation within the single Mie-elements (quasi BIC-state) with the overall radiation loss cancellation of the chain. From this a further substantial enhancement of light-matter interaction is expected.  

\subsection{Conclusion}

To conclude, we have designed and fabricated optical cavities based on  linear arrays of coupled silicon pillars. The coupling between resonant Mie modes of each pillar results in the formation of collective resonant states with suppressed radiation losses and increased Q-factor. We have experimentally demonstrated that these modes allow for the enhancement of the light emission from Ge(Si) quantum dots embedded inside the pillars and that the Q-factor of the the collective Mie resonances increases with the length of the chain. The  measured Q-factor of a collective magnetic mode for an array of 11 pillars as high as $Q\approx 500$ was demonstrated, which was supported by numerical simulations.  
We showed the connection of the considered high-Q modes and the modes of an infinite discrete waveguide located at the band edge (flat band region). 
Thus, with the coupled Mie-resonators, one can realize a bottom-up approach to designing the Bloch-Floquet optical cavities contrary to the common top-down strategy of designing photonic crystal cavities. In the proposed approach, embedding quantum emitters inside coupled Mie-resonators opens the potential of engineering optical cavities for active photonic devices with the footprint smaller than alternative microresonant systems.

\section{Methods}

{\it Fabrication.} The pillars of the proposed geometry were fabricated within the common planar technology. At the first step, the structures with Ge(Si) self-assembled QDs were grown by molecular beam epitaxy at 650$^0$ C. For this, a commercial silicon-on-insulator (SOI) substrate with 3 $\mu$m buried oxide (BOX) layer and a thinned down 90 nm Si device layer was used as substrate. The grown structures consist of 75 nm Si buffer layer, lattice with 12 layers of Ge(Si) QDs separated by 15 nm Si spacers. The growth was finalized by the deposition of a 165 nm Si capping layer. Growth details and QD parameters can be found in Ref.~\citenum{NOVIKOV2004416}. The total thickness of the structure above the BOX amounts to 500 nm.

Single pillar and chain resonators were fabricated from the epitaxial structure using electron-beam lithography and inductively coupled plasma reactive ion etching (ICP-RIE). For this, a 30 nm Cr hard mask was deposited; a 200 nm thick negative resist was spin-coated on top prior to the electron beam exposure. Subsequently the Cr mask was etched using Cl$_2$ and O$_2$ gases, and finally the Si device layer was etched using SF$_6$ and C$_4$F$_8$ gases.

{\it Optical characterization} Optical properties of fabricated resonators were investigated using micro-photoluminescence measurements (\textit{Horiba LabRAM setup}). A 532 nm continuous wave laser was focused on the sample from the top using a 100X objective lens (\textit{Mitutoyo LD APO NIR}, NA=0.9) resulting in a spot size of about 1.5 $\mu$m on the sample. The emitted light was collected by the same objective and spectrally resolved by a 150 or 600 lines per mm grating (blazed at 1200 and 750 nm, respectively) in a monochromator. After that, the light is detected by a liquid nitrogen-cooled InGaAs CCD camera (\textit{Symphony II}). The setup is equipped with an adjustable mirror, which can deflect the excitation laser beam, thus allowing the PL mapping of the sample with a spatial resolution of 100 nm. For the polarization measurements, a polarizer was inserted in the optical path before the monochromator.

{\it Numerical modeling} Numerical modeling of the eigenmodes of an isolated pillar and finite chains was performed in commercially available finite-element based software COMSOL Multiphysics, using the Wave optics module and taking into account the \sio substrate and Si dispersion. Dispersion diagrams of infinitely extended arrays were obtained in CST Microwave Studio for the same parameters as finite arrays.

\begin{acknowledgement}

The authors thank Sven Schlenker for the help in sample fabrication.

M.P. acknowledges support from DAAD, grant \# 57387479. 

V.R. and J.S. thank the Federal Ministry for Education and Research (“Bundesministerium für Bildung und Forschung”, Project N. 03Z2HN12) for their financial support within the Centre for Innovation Competence SiLi-Nano\textsuperscript{\textregistered}.

This project has received funding from Joint German-Russian Research Project RFBR-DFG \# 20-52-12062.

This project has received funding from the European Union’s Horizon 2020 research and innovation
programme under the Marie Sklodowska-Curie grant agreement No 845287.

The authors declare no competing financial interest.

\end{acknowledgement}

\begin{suppinfo}

Polarization-resolved $\mu$PL from unprocessed layers, PL map of the chain, Parameters variation of the chain.

\end{suppinfo}

\bibliography{Main}

\newpage
\begin{figure}[t]
\includegraphics{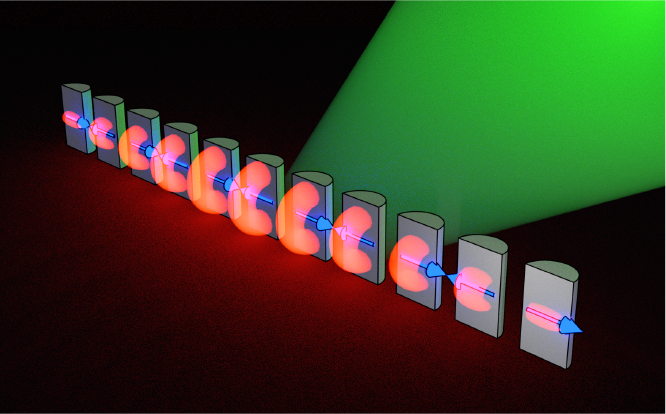}
\caption{TOC}\label{TOC}
\end{figure}

\end{document}